# A Safe DRL Method for Fast Solution of Real-Time Optimal Power Flow

Pengfei Wu[a,b], Chen Chen[a,b,*], Dexiang Lai[a,b], and Jian Zhong[a,b]

[a]*State Key Laboratory of Electrical Insulation and Power Equipment, Xi'an Jiaotong University, Xi'an, 710049, Shaanxi, China*
[b]*Shaanxi Key Laboratory of Smart Grid, Xi'an Jiaotong University, Xi'an, 710049, Shaanxi, China*

**Abstract**

High-level penetration of intermittent renewable energy sources (RESs) has introduced significant uncertainties into modern power systems. In order to rapidly and economically respond to the fluctuations of power system operating state, this paper proposes a safe deep reinforcement learning (SDRL) based method for fast solution of real-time optimal power flow (RT-OPF) problems. The proposed method considers the volatility of RESs and temporal constraints, and formulates the RT-OPF as a Constrained Markov Decision Process (CMDP). In the training process, the proposed method hybridizes the proximal policy optimization (PPO) and the primal-dual method. Instead of integrating the constraint violation penalty with the reward function, its actor gradients are estimated by a Lagrange advantage function which is derived from two critic systems based on economic reward and violation cost. The decoupling of reward and cost alleviates reward sparsity while improving critic approximation accuracy. Moreover, the introduction of Lagrange multipliers enables the agent to comprehend the trade-off between optimality and feasibility. Numerical tests are carried out and compared with penalty-based DRL methods on the IEEE 9-bus, 30-bus, and 118-bus test systems. The results show that the well-trained SDRL agent can significantly improve the computation efficiency while satisfying the security constraints and optimality requirements.

*Keywords:* Real-time optimal power flow, safe deep reinforcement learning, primal-dual proximal policy optimization.

## 1. Introduction

Serving as the foundation for power system planning, operation and market analysis, the optimal power flow (OPF) is a crucial aspect of power system research. The OPF helps to determine the optimal generator set-points to minimize operational cost, voltage deviation, network power loss, and other objectives while satisfying security constraints [1]. With the increasing penetration of intermittent energy resources, there are two severe challenges for the OPF problem: 1) It is a large-scale non-convex constrained optimization problem, which has been proven to be an NP-hard problem [2]; 2) Due to the rapid power flow fluctuations caused by the integration of renewable energy sources (RESs) in the modern power system, timely control commands are required for the secure and economic system operation, which requires limiting the solving time for AC OPF. The day-ahead scheduling or the traditional OPF with slow timescale (say every tens of minutes) may not be suitable for future smart grids. To describe the demand of solving the OPF in time, the concept of real-time OPF (RT-OPF) has been proposed in [3]. RT-OPF requires solving the OPF on a faster timescale, to ensure that the operation scheme could track the time-varying demands and renewable generations. Therefore, enhancing the computation efficiency of OPF problems while ensuring (near) optimality has been a hotspot of research in recent years.

There have been many mathematical optimization methods to tackle the AC OPF problem. The simplified gradient method is proposed by Dommel and Tinney in [4], which is a classical algorithm to solve the large-scale OPF problem. Because of its high efficiency and numerical stability, interior-point method is widely used to solve the AC OPF problem [5]-[6], and the optimal solution could be found by some popular nonlinear solvers, e.g. IPOPT [7]. The convex relaxation method has been applied to solve the AC OPF problems through second-order cone programming (SOCP) [8] and semidefinite programming (SDP) [9]. A sufficient condition for recovering the optimal solution from the relaxation solution is presented in [10]. In addition, linear approximation is also considered to solve the AC OPF problem. In [11], a detailed analysis and breakdown investigation of existing linear approximations for OPF problem are provided and the accuracy of existing linear approximations is compared. In [12], a linear relaxation-based SOCP method for bi-objective DC OPF is presented. The aforementioned methods are based on the mathematical model that boasts high interpretability. However, this kind of methods usually suffers from huge computational burden and requires a considerable time to achieve convergence. For modern power systems containing more volatility and uncertainty, it is difficult to meet the solving speed requirements of real-time operation, which limits the practical application of these methods.

To address this issue and achieve efficient solutions for power system operation problems, data-driven approach based on machine learning (ML) is considered as one of the most promising approaches [14]. It shifts the computational burden to the offline training process, improving the computation speed. There has been a lot of research work on OPF utilizing deep neural networks (DNN) and supervised learning techniques [15]-[20]. The DNN is utilized to approximate the solutions to DC and AC OPF in [15] and [16]. In [17], inactive constraint sets are classified by the DNN under different demand conditions, which reduces the computational burden. In [18], the voltage magnitudes and phase angles of the OPF solution are estimated by DNN, then the set-points of generators are obtained from the power flow balance equation. In [19], a physics-informed neural network (PINN) is proposed for the AC OPF, which can improve the prediction accuracy. In [20], the major research work in this area is summarized and a concise framework for generalizing ML-based methods is introduced. The performance of fully connected (FCNN), convolutional (CNN), and graph (GNN) neural networks is also compared in [20]. These methods use supervised learning techniques to obtain the optimal solution or reduce the computational burden of the OPF problems, which can achieve solution acceleration. However, these methods still suffer from generalization errors and iterative processes, which may make such methods fail to fully guarantee the system security constraints with weak credibility or insignificant computational speedup.

In addition to supervised learning, another promising class of data-driven methods is deep reinforcement learning (DRL) technique [21], which has shown great potential in solving complex decision-making problems through trial-and-error interactions with the environment. A well-trained DRL agent can make decisions based on necessary observations without relying on a sophisticated mathematical model, thus enabling instantaneous response to varying environmental conditions. Relevant studies have been conducted in the research field of DRL-based applications for smart grid, including economic dispatch [22], voltage regulation [23], active distribution system operation [24], and frequency control [25]. For the AC OPF problem, some classical DRL algorithms are also applied to achieve computation acceleration. In [26], twin delayed deep deterministic policy gradient (TD3) algorithm is employed to regulate the generator set-points with the aim of minimizing the power system operational cost. Proximal policy optimization (PPO) algorithm is utilized to tackle the AC OPF problem and the robustness of PPO is demonstrated in [27]. In [13] and [28], a Lagrangian based deep deterministic policy gradient (DDPG) approach is applied to the AC OPF and security-constrained optimal power flow (SCOPF) problems. The critic network is replaced by a Lagrangian function and the interior-point method is used to directly estimate the gradients of the actor network, which stabilizes the training process and enhances convergence.

Compared to the supervised learning, DRL methods are capable of addressing sequential decision problems and enhancing model interpretability through the construction of reward function [13]. However, most studies choose to incorporate the penalty terms into the reward function to account for the security constraints in power system problems [22]-[27]. Determining the optimal coefficients for the penalty term of reward function poses a challenge in hyper-parameter selection, i.e., too large coefficients will result in an excessive gap between the objective and reward function, while too small coefficients may lead to weak binding capacity. Additionally, this method increases the reward function sparsity and reduces the accuracy of critic network estimates, making it difficult for the agent to fully ensure the operational security [28]. The primary reason lies in the insufficient refinement of Markov Decision Process (MDP) for constraints, which fails to effectively reflect the priority between feasibility and optimality. As a result, the feasibility premise is easily overlooked in pursuit of the highest reward.

In this paper, we focus on how to fully consider system operational constraints during training process and ensure the priority of feasibility. Inspired by the research of safe DRL (SDRL) in [29] and [30], we propose a new method for solving RT-OPF problems, which combines model-free DRL algorithms with the mathematical optimization methods to achieve the balance of optimality and feasibility during the training process. More specifically, the significant contributions of this paper are outlined below.

First, the real-time OPF problem is formulated as a Constrained Markov Decision Process (CMDP), where time-dependent constraints are considered. By directly modelling the operational constraints, the agent has access to security-related gradient information for seeking feasible region during the training process.

Second, a safe Actor-Critic DRL algorithm, namely primal-dual proximal policy optimization (PD-PPO) algorithm, is proposed to solve the formulated CMDP. More specifically, the fundamental algorithm PPO is extended to the CMDP considering the characteristics of the real-time AC OPF problem. The application of the PD-PPO algorithm enables agents to adaptively balance the feasibility and optimality of learning objectives during the training process. In this way, the agent prioritizes constraint security as primary objective and conducts optimization to search within the feasible region.

Third, a DRL training method coordinated with imitation learning (IL) technique and expert trajectories is constructed for solving the RT-OPF problems. The IL is utilized to pre-train agents to reduce the exploration space. Each single step of the expert trajectory is expanded into a training episode for better convergence, while the entire scheduling cycle is considered as an episode during the testing process. This method utilizes the prior knowledge, mitigates the environmental randomness, reduces the episode length, and promotes the convergence of the critic network.

Compared with the supervised learning and penalty-based DRL methods, the method proposed in this paper is more interpretable due to the decoupling of objective and constraint modelling. The proposed PD-PPO method boasts several advantages, including high training efficiency, decreased computational complexity of AC OPF, strict adherence to security constraints, and near-optimality of solution. The effect of the proposed method is proved through case studies and compared with two types of existing penalty-based DRL methods on three IEEE test systems.

The remainder of the paper is organized as follows. Section 2 provides the problem formulation of AC OPF and the principles of safe DRL. Section 3 details the CMDP model of the RT-OPF problem and the training process for agents. The proposed PD-PPO methodology is presented in Section 4. In Section 5, case studies are conducted using the IEEE 9-bus, IEEE 30-bus, and IEEE 118-bus to demonstrate the effectiveness of the proposed method. Conclusions and future work of this paper are presented in Section 6.

## 2. Problem Formulation and Safe DRL

*A. Definition of the real-time OPF problem*

We consider a power system consisting of $N$ nodes denoted by the set of buses $\mathcal{N}$, and $L$ transmission branches by the set of branches $\mathcal{L}$, where generator buses are sorted out in a subset $\mathcal{G} \subset \mathcal{N}$. The objective of the RT-OPF problem is to find the optimal set-points of the generators that minimize the operating cost and satisfy all the operational constraints of the power system for each dispatch interval *t*. Different from the traditional OPF, the real-time OPF should find the optimal operating state on a faster timescale in response to the highly rapid and uncertain fluctuations in both generation and load sides. The mathematical model of the RT-OPF problem can be formulated as [3]:

$$\min \sum_{i \in \mathcal{G}} \left( c_{2i} P_{gi}^2 + c_{1i} P_{gi} + c_{0i} \right), \tag{1}$$

s.t.

$$P_{gi}^{\min} \leq P_{gi} \leq P_{gi}^{\max}, \forall i \in \mathcal{G}, \tag{2}$$

$$Q_{gi}^{\min} \leq Q_{gi} \leq Q_{gi}^{\max}, \forall i \in \mathcal{G}, \tag{3}$$

$$V_i^{\min} \leq |V_i| \leq V_i^{\max}, \forall i \in \mathcal{N}, \tag{4}$$

$$P_{flow,ij}^2 + Q_{flow,ij}^2 \leq \left| S_{flow,ij}^{\max} \right|^2, \forall (i,j) \in \mathcal{L}, \tag{5}$$

$$P_{flow,ij} = g_{ij}^{ch} V_i^2 + V_i V_j \left[ b_{ij} \sin \delta_{ij} - g_{ij} \cos \delta_{ij} \right], \forall (i,j) \in \mathcal{L}, \tag{6}$$

$$Q_{flow,ij} = b_{ij}^{ch} V_i^2 + V_i V_j \left[ g_{ij} \sin \delta_{ij} + b_{ij} \cos \delta_{ij} \right], \forall (i,j) \in \mathcal{L}, \tag{7}$$

$$P_{gi} - P_{di} = |V_i| \sum_{j \in i} |V_j| \left( g_{ij} \cos \delta_{ij} + b_{ij} \sin \delta_{ij} \right), \forall i \in \mathcal{N}, \tag{8}$$

$$Q_{gi} - Q_{di} = |V_i| \sum_{j \in i} |V_j| \left( g_{ij} \sin \delta_{ij} - b_{ij} \cos \delta_{ij} \right), \forall i \in \mathcal{N}, \tag{9}$$

$$-R_{gi}^{down} \leq P_{gi}^t - P_{gi}^{t-1} \leq R_{gi}^{up}, \forall i \in \mathcal{G}, \tag{10}$$

where, $P, Q, V$ are the active power, reactive power, and bus voltage amplitude, respectively; $\delta_{ij} = \alpha_i - \alpha_j$ is the voltage phase angle difference between the bus *i* with bus *j*; $c_{0i}$, $c_{1i}$, and $c_{2i}$ are the coefficients of the quadratic costs of the generator *i*; the subscripts *g*, *d*, and *flow* represent the generator, demand, and transmission line, respectively; the superscripts min and max represent the upper and lower bounds of the constraints; $g_{ij}/b_{ij}$ and $g_{ij}^{ch}/b_{ij}^{ch}$ are the series and shunt conductance/susceptance of the transmission line between bus *i* and bus *j*; $R_{gi}^{down}/R_{gi}^{up}$ is the maximum ramping down/up rate. Constraints (2)-(7) denote the limitation of active/reactive power output of generators, voltage amplitude, and transmission capacity. Constraints (8) and (9) are the power flow equations. Constraint (10) is incorporated into the AC OPF problem to describe the temporal constrains about ramping capabilities. The objective of this model is then to find a high-quality near optimal policy for the time-varying RT-OPF, given current states of power system and previous generator set-points [3].

*B. Safe DRL Basis*

Safe deep reinforcement learning is an extension of the DRL, which additionally utilizes a cost function to consider the policy security. Based on a Constrained Markov Decision Process, the objective of the SDRL is maximizing the long-term expected reward while satisfying constraints on the long-term expected cost [29]. CMDP can be represented by a tuple $(\mathcal{S}, \mathcal{A}, P, R, C, \gamma)$, where $\mathcal{S}$ denotes the set of states; $\mathcal{A}$ represents the set of actions; $P: \mathcal{S} \times \mathcal{A} \times \mathcal{S} \mapsto [0,1]$ refers to the environment transition probability function; $R: \mathcal{S} \times \mathcal{A} \times \mathcal{S} \mapsto \mathbb{R}$ stands for the reward function; $C_i: \mathcal{S} \times \mathcal{A} \times \mathcal{S} \mapsto \mathbb{R}$ is the cost vector function whose corresponding limit is $d_i$; $\gamma$ is a discount factor ($0 \leq \gamma \leq 1$), which determines the weight given to future rewards versus immediate rewards during the decision-making process. For a stochastic policy $\pi: \mathcal{S} \mapsto \mathcal{P}(\mathcal{A})$, it corresponds to a mapping from states to a probability distribution over actions. Specifically, $\pi(a|s)$ denotes the probability of selecting action *a* in state *s* [21].

At each time step *t*, the agent selects an action $a_t$ based on the current environment state $s_t$, then the environment transitions to the next state $s_{t+1}$ according to the transition probability $P(s_{t+1}|s_t, a_t)$ and feedbacks a reward $r_t = R(s_t, a_t, s_{t+1})$ and a cost vector $c_t = C(s_t, a_t, s_{t+1})$ to the agent. After the agent interacts with the environment, a set of interaction trajectories are obtained, which can be denoted as $\tau = (s_0, a_0, s_1, a_1, ...)$. The discounted reward and cost can be calculated by (11) and (12). The long-term expected reward and cost is defined as (13) and (14):

$$G_t^R = \sum_{k=0}^{\infty} \gamma^k R(s_{t+k}, a_{t+k}, s_{t+k+1}), \tag{11}$$

$$G_{i,t}^C = \sum_{k=0}^{\infty} \gamma^k C_i(s_{t+k}, a_{t+k}, s_{t+k+1}), \tag{12}$$

$$R(\pi) = \mathbb{E}_{a_t \sim \pi(\cdot|s_t), s_{t+1} \sim P(\cdot|s_t, a_t)} \left[ \sum_{t=0}^{\infty} \gamma^t R(s_t, a_t, s_{t+1}) \right], \tag{13}$$

$$C_i(\pi) = \mathbb{E}_{a_t \sim \pi(\cdot|s_t), s_{t+1} \sim P(\cdot|s_t, a_t)} \left[ \sum_{t=0}^{\infty} \gamma^t C_i(s_t, a_t, s_{t+1}) \right], \forall i \in [m], \tag{14}$$

For the safe DRL, the policy is determined by an actor DNN with parameters $\theta$ to be updated. The mathematical formulation to optimize the parameters of the actor DNN can be presented as follows:

$$\pi^* = \arg\max_{\theta} R(\pi_\theta),$$
$$s.t.\ C_i(\pi_\theta) \le d_i,\ \forall i \in [m]. \tag{15}$$

Compared to the classical MDP, the CMDP is obviously more suitable for the constrained optimization problems such as RT-OPF. The reward and cost function can correspond to the objective function and security operational constraints, which facilitates more flexible design of training algorithms.

## 3. CMDP Formulation and Training Process

### A. CMDP formulation of the real-time OPF

In the RT-OPF, the power system operator can be viewed as an agent to interact with the power system through selecting actions to minimize the operational cost while considering the security constraints. Fig. 1 illustrates the learning structure and interaction process between the agent and the power system environment within a CMDP model. Based on the interaction trajectory and the reward-cost sequence, the PD-PPO algorithm iteratively updates the control policy to achieve the performance improvement.

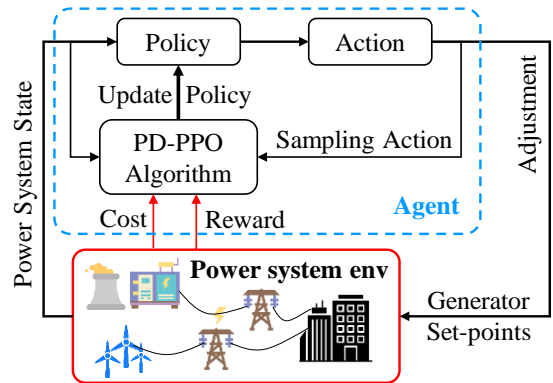

***Fig. 1.*** *Diagram of the interaction process between agent and environment.*

To reformulate the RT-OPF problem as a sequential decision-making problem, the CMDP is represented by five critical elements $(\mathcal{S}, \mathcal{A}, P, R, C)$, which are defined as follows.

*1) State Space $\mathcal{S}$*: The observed states should be able to describe the current environment sufficiently to enable agents to make accurate decisions. Considering the uncertainty of power demands and renewable energy within power system, the net load of each bus is included in the state space. To handle the time-dependent constraints (10), the state space should include two other terms: the set-points of generators at previous time and the sum of net load at next time. Consequently, the observed states space is finally determined to include:

$$\mathcal{S} = \left\{ s_t := \left( P_i^t, Q_i^t, \forall i \in \tilde{\mathcal{N}}, P_{gi}^{t-1}, V_{gi}^{t-1}, \forall i \in \mathcal{G}, \sum_{i \in \tilde{\mathcal{N}}} P_i^{t+1} \right) \right\}, \tag{16}$$

where $\tilde{\mathcal{N}} \subset \mathcal{N}$ is a subset where net load is nonzero; $P_i^t$ and $Q_i^t$ are the active and reactive net load of bus $i$ at current time $t$; $P_{gi}^{t-1}$ and $V_{gi}^{t-1}$ are the set-points of generator $i$ at previous time; $\sum_{i \in \tilde{\mathcal{N}}} P_i^{t+1}$ is the estimation of the net load for whole power system at next time. The dimension of the state space is $2|\tilde{\mathcal{N}}| + 2|\mathcal{G}| + 1$.

*2) Action Space $\mathcal{A}$*: Minimizing operational cost can be achieved by controlling the active output and voltage set-points of generators. To reflect the temporal relationships during the training process and correlate state and action spaces, the action space includes the adjustments of active output and voltage set-points of generators, which is defined in (17):

$$\mathcal{A} = \left\{ a_t := \left( \Delta P_{gi}^t, \Delta V_{gi}^t, \forall i \in \mathcal{G} \right) \right\}. \tag{17}$$

Then the current set-points of generators can be calculated through $P_{gi}^t = P_{gi}^{t-1} + \Delta P_{gi}^t$ and $V_{gi}^t = V_{gi}^{t-1} + \Delta V_{gi}^t$. The dimension of the action space is $2|\mathcal{G}|$.

*3) Transition Probability $P$*: After the agent selects an action $a_t$, the environment will transfer from state $s_t$ to a new state $s_{t+1}$ with the transition probability $P(s_{t+1}|s_t, a_t)$. Due to the randomness of the real world, the transition probability usually cannot be expressed formulaically, such as the load demand and renewable energy generation. To handle the environment with transition uncertainty, a model-free DRL algorithm and an imitation-based training method are proposed in this paper, as discussed in Section 4.

*4) Reward Function $R$*: The reward function with penalty terms will cause difficulty in parameter selection and negatively impact the convergence of the critic network, which further limits the training efficiency of the DRL algorithm. In this paper, the CMDP model decouples the reward and cost function for a more precise characterization of objective and constraints, which facilitates stable and efficient training. Considering the objective function of RT-OPF (1), the reward function should be negatively correlated with the system operational cost. The specific form is as follows:

$$reward = -\sum_{i \in \mathcal{G}} \left( c_{2i} P_{gi}^2 + c_{1i} P_{gi} + c_{0i} \right). \tag{18}$$

Unlike the reward function designed in [22]-[27], the reward function (18) is very smooth and doesn't contain any penalty terms, so the critic network with a simple structure could achieve convergence efficiently.

    5) *Cost Function $C$*: Similar to the reward function, the penalty function is positively correlated with system constraint violation, which can be expressed as:

$$Cost = \left[ C_{pg}, C_{qg}, C_{vg}, C_{flow} \right], \tag{19}$$

where $C_{pg}$, $C_{qg}$, $C_{vg}$, and $C_{flow}$ are the violation of active/reactive power constraints, bus voltage amplitude constraints, and transmission capacity constraints. In particular, the cost function is a vector function unlike the reward function, which has good scalability. If the SCOPF problem is considered, the cost function can be further expanded as a vector with $4(N_C+1)$ elements:

$$Cost_{SC} = \left[ Cost_0, \cdots, Cost_i, \cdots, Cost_{N_C} \right], i = 0, ..., N_C, \tag{20}$$

where $N_C$ is the number of contingencies considered and the contingency index $i=0$ corresponds to the base case. When dealing with the SCOPF problem, expanding for critic networks is sufficient because the constraint violations under different contingency cases can be obtained easily, thus ensuring the scalability of the proposed method.

*B. Offline environment construction*

    The agent is trained offline through interactions with a simulated environment before online applications in a real-world power system. The offline simulated environment in this paper is established using OpenAI Gym [31] and PYPOWER (which is the python version of MATPOWER [32]). Gym is the benchmark systems for DRL studies, which is developed by OpenAI. PYPOWER provides the Newton-Raphson Power Flow solver and interior-point solver (IPS), which enables researchers to construct a training environment more conveniently. The training dataset is generated by applying certain random perturbations to the real-world system data. Supposing the training dataset contains demand conditions for *T* steps, the initialization system parameters are generated as follows:

$$P_i^t \sim \mathbb{U}\left( \underline{P}_i^t, \overline{P}_i^t \right), t \in [1,T], i \in \tilde{\mathcal{N}}, \tag{21}$$

$$\beta_i^t \sim \mathbb{U}\left( \underline{\beta}_i^t, \overline{\beta}_i^t \right), t \in [1,T], i \in \tilde{\mathcal{N}}, \tag{22}$$

$$Q_i^t = P_i^t \tan\left( \cos^{-1}\left( \beta_i^t \right) \right), t \in [1,T], i \in \tilde{\mathcal{N}}, \tag{23}$$

$$P_g^0 = runopf\left( P_i^{t-1}, Q_i^{t-1} \right), t \in [2,T], i \in \tilde{\mathcal{N}}, \tag{24}$$

where $\mathbb{U}$ means uniform distribution and $\beta$ is power factor which is calculated from the base case of PYPOWER. It is worth mentioning that the initial set-points of the generators are optimized by the IPS based on the demand conditions of the previous *t*-1 time, which is one of the bases of the imitation-based training approach.

    For the RT-OPF problem, we expect a well-trained agent that could achieve the optimal or near-optimal operation of the power system by single-step adjustment of set-points in real time. To achieve efficient training for the RT-OPF, the problem is decomposed and modelled as a series of single-step sub-problems within a scheduling cycle, while each sub-problem considers time-dependent constraints to guarantee the equivalence with the original problem. Then the single-step decision problem is expanded into an episode for agent training to reduce the complexity resulted from the environment randomness.

    The PD-PPO agent could imitate the optimal decisions obtained by the interior-point solver to accelerate convergence. In order to collect the prior knowledge for imitation, the interior-point solver is considered as an expert policy and utilized to generate a series of interaction trajectories based on the training dataset. Then, each step of the trajectory is treated as an episode of the DRL training, where the demand conditions remain constant and the agent explores through trial and error to find the optimal action. This concept is illustrated in Fig. 2 below.

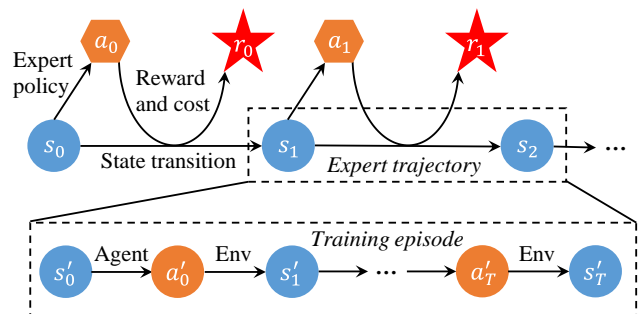

**Fig. 2.** *Offline training dataset construction.*

Based on the trajectory data of the expert policy, the specific interaction flow chart of the training process is shown in Fig. 3 below. At the beginning of each episode, the $Reset(\cdot)$ function is first called to sample a time step from the expert trajectory as the target problem. Then the interaction between agent and environment is carried out by calling the $Step(\cdot)$ function. When the maximum number of steps in the episode is reached, the episode is over and the next episode is ready to begin. The optimal policy is expected to directly adjust generators to their optimal set-points at the first step of each episode, with no further adjustment in subsequent steps. It achieves the same characteristics as the real-time response optimal policy in real-world application. The idea of imitation-based methods is similar to the principle of the "target" critic network in the DDPG algorithm because it is always easier to solve a "fixed target" than a "moving target".

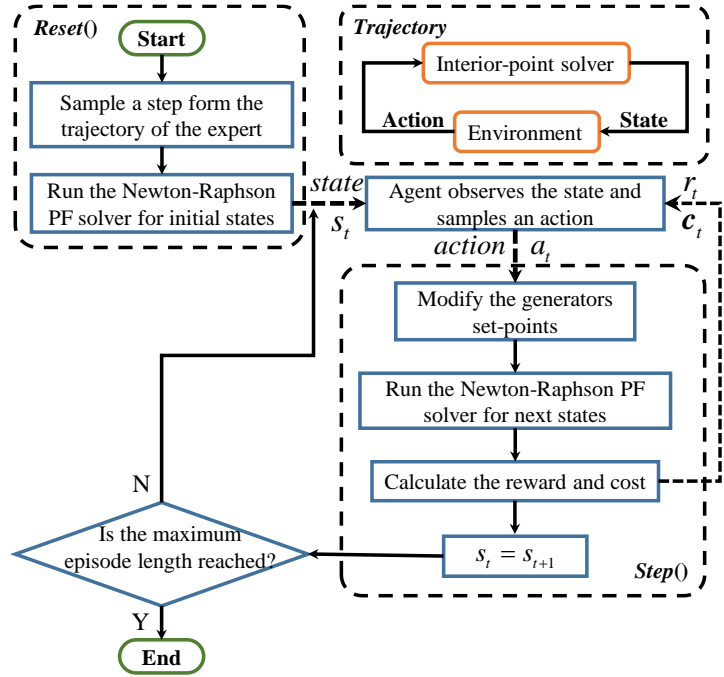

*Fig. 3. Diagram of the training process.*

## 4. Primal-dual PPO Algorithm

The model-free DRL approach has demonstrated excellent performance on many real-world challenging tasks, such as video games, robotics, and automatic driving [27]. As the extension of the policy-based algorithm and one of state-of-the-art DRL algorithms, the PPO [34] algorithm has good convergence property and generalizability on different tasks, which is considered as the default algorithm by OpenAI. Additionally, the PPO algorithm achieves stable policy improvement by updating in the "trust region", and is insensitive to the problem dimension and parameter settings, which is more suitable to deal with the RT-OPF problems in this paper. In this section, we develop a primal-dual PPO (PD-PPO) algorithm to handle the formulated CMDP of the RT-OPF problem while ensuring secure control actions. The training techniques of the PD-PPO and implementation details are represented in the following subsections.

### A. PD-PPO algorithm

The advantage of the PPO is derived from a series of training techniques, including actor-critic structure, generalized advantage estimation (GAE) [35], importance sampling (IS), and gradient-clip. The above techniques can also be extended to the CMDP problem described in this paper. Therefore, the primal-dual PPO algorithm for solving CMDP can be constructed based on these techniques.

For the proposed PD-PPO, a Lagrangian function is applied to convert the constrained problem (15) to the unconstrained problem:

$$L(\pi_\theta, \lambda) = R(\pi_\theta) - \sum_i^m \lambda_i \left( C_i(\pi_\theta) - d_i \right),$$
$$(\pi^*, \lambda^*) = \arg\min_{\lambda_i \geq 0} \max_\theta L(\pi_\theta, \lambda), \tag{25}$$

where $L(\pi_\theta, \lambda)$ represents the augmented objective function containing the reward function and cost function; $\lambda_i$ denotes the Lagrangian multiplier of the *i*th constraint. To solve the training problem (25), the proposed PD-PPO algorithm utilizes two critic networks to evaluate the performance of the actor network in terms of feasibility and optimality respectively, then optimizes the parameters of the actor network based on the following training techniques. The detailed workflow of the proposed PD-PPO algorithm is shown in Fig. 4 below.

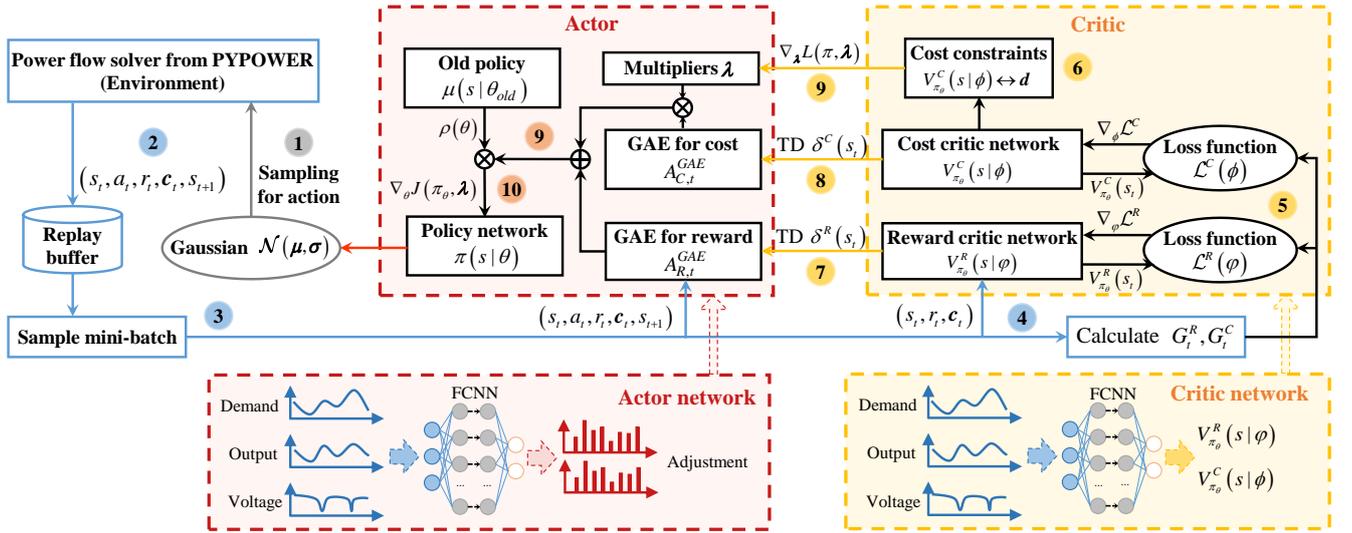

**Fig. 4.** *Workflow of the proposed PD-PPO methodology.*

*1) Generalized Advantage Estimation:* In PPO, the GAE function is used to evaluate the advantage of the selected action relative to a random action, which can effectively mitigate variance during the training process [35]. For the GAE $A_{R,t}^{GAE}$ of reward, the calculation method is shown below:

$$\hat{A}_{R,t}^{(k)} = \sum_{l=0}^{k-1} \gamma^l \delta_{t+l}^R, \tag{26}$$

$$\delta_t^R = r_t + \gamma V_\pi^R(s_{t+1}) - V_\pi^R(s_t), \tag{27}$$

$$A_{R,t}^{GAE} = (1-\lambda_{GAE})\left(\hat{A}_{R,t}^{(1)} + \lambda_{GAE}\hat{A}_{R,t}^{(2)} + \lambda_{GAE}^2 \hat{A}_{R,t}^{(3)} + \cdots\right), \tag{28}$$

where $\hat{A}_{R,t}^{(k)}$ is $k$-step advantage estimation; $\delta_t^R$ represents the temporal difference (TD) residual of $V_\pi^R(s)$; $V_\pi^R(s)$ is the state value function that measures the quality of state $s$ under the policy $\pi$, which is the output of the reward critic network; $\lambda_{GAE}$ is a parameter which controls the average degree of $n$-step advantage values. For the cost function, it is similar to derive the corresponding GAE function $A_{C,t}^{GAE}$. Different from the PPO algorithm in [34], the GAE function for problem (25) can be obtained as:

$$A_{L,t}^{GAE} = A_{R,t}^{GAE} - \sum_{i}^{m} \lambda_i A_{C_i,t}^{GAE}. \tag{29}$$

It should be emphasized that the Actor-Critic architecture guarantees convergence improvement only under certain conditions, i.e., the accuracy of the critic network satisfies the requirement of policy update. For the PPO algorithm, the accuracy requirement of the critic network can be relaxed because the gradient for policy originates from the smoother GAE function. Therefore, the dependence between the actor and critic is weakened, which contributes to the stability of algorithm convergence.

*2) Importance Sampling:* The PPO algorithm applies the importance sampling during the training process for higher sample efficiency [34]. Based on the training data obtained by the old policy $\pi_{\theta'}$, the objective function of the updated policy $\pi_\theta$ is:

$$J^{\theta'}(\theta) = E_{(s_t,a_t)\sim\pi_{\theta'}}\left[\frac{P_\theta(a_t|s_t)}{P_{\theta'}(a_t|s_t)} A_{L,t}^{\theta'}(s_t,a_t)\right], \tag{30}$$

where $(s_t,a_t)\sim\pi_{\theta'}$ represents the batch of training data from the interaction trajectory between the old policy $\pi_{\theta'}$ and environment; $A_{L,t}^{\theta'}(s_t,a_t)$ denotes the Lagrangian GAE (29) of training data generated by $\pi_{\theta'}$.

The prerequisite for the importance sampling is that the updated policy $\pi_\theta$ must not differ too much from the old policy $\pi_{\theta'}$. Since $\pi_\theta$ in PPO is a random policy, measuring the difference between $\pi_\theta$ and $\pi_{\theta'}$ can be achieved by calculating the KL-Divergence:

$$D_{KL} = \sum_{(s_t,a_t)} P_\theta(a_t|s_t)\left(\log P_\theta(a_t|s_t) - \log P_{\theta'}(a_t|s_t)\right). \tag{31}$$

When training with the same batch of interaction data, the training process should be stopped if the KL-Divergence is higher than a threshold $KL_{tar}$. Then a new batch of training data needs to be generated for training to ensure the accuracy of the importance sampling.

*3) Gradient-Clip technique:* The PPO updates DNN parameters within a "trust region" to realize the stable improvement of policy performance [36]. Checking the KL-Divergence during the training process can constrain the policy update. To further regulate the update process, the gradient of objective function can be clipped as the below equation:

$$J(\theta) = E_{(s_t,a_t) \sim \pi_{\theta'}} \left[ \min\left(\rho(\theta) A_{L,t}^{\theta'}, clip(\rho(\theta), 1-\varepsilon, 1+\varepsilon) A_{L,t}^{\theta'} \right) \right], \tag{32}$$

where $\rho(\theta) = p_\theta(a_t | s_t)/p_{\theta'}(a_t | s_t)$; $A_{L,t}^{\theta'}$ is the augmented GAE function (29); $clip(\cdot)$ function is utilized to constrain $\rho(\theta)$ within a certain range $[1-\varepsilon, 1+\varepsilon]$, thereby preventing excessive policy updates caused by too large gradient.

For complex tasks, the value function hyperplane is often highly complex, making it difficult to determine an appropriate learning rate. During the training process, there is a risk of falling off the "cliff" of the hyperplane, which can result in fluctuations of policy performance. However, the PPO policy updates are limited within a "trust region" through the KL-divergence constraint and Gradient-Clip technique, thus ensuring the monotonicity of the policy improvement.

*4) Update dual multipliers and Critic DNN:* Similar to the primal-dual approach, the PD-PPO algorithm updates the actor network and dual multipliers in turn. According to the gradient descent, the gradient of the dual multiplier update is shown below.

$$\begin{cases} \nabla_\lambda L(\pi_\theta, \lambda) = \frac{1}{N_{batch}} \sum_{s_t \in D_{batch}} \left[ \rho(\theta) V_\pi^C(s_t) - d \right]^+ \\ \lambda = \left[ \lambda + \eta_\lambda \nabla_\lambda L(\pi, \lambda) \right]^+ \end{cases}, \tag{33}$$

where $N_{batch}$ is the capacity of PD-PPO replay buffer $D_{batch}$; $[x]^+$ means $\max(x, 0)$; $V_\pi^C(s_t)$ refers to the output of the cost critic network.

If the security constraints are not satisfied under the policy $\pi$, the value function $V_\pi^C(s_t)$ will exceed the limits and the corresponding Lagrange multiplier will improve, making the agent prioritize the policy security. After adequate training, the security of the policy will improve and then the multipliers begin to remain constants. In this case, the training objective shifts to minimizing objective function gradually. The agent will balance the feasibility and optimality based on Lagrange multipliers during the training process and prioritize the security of constraints.

*5) Update Critic networks:* The update of the critic networks is achieved by Monte Carlo sampling. The loss functions of the reward and cost critic network are shown as follows.

$$L_{critic}^R = \frac{1}{N_{batch}} \sum_{(s_t, G_t^R) \in D_{batch}} \left[ G_t^R - V_\pi^R(s_t) \right]^2, \tag{34}$$

$$L_{critic}^C = \frac{1}{N_{batch}} \sum_{(s_t, G_t^C) \in D_{batch}} \left[ G_t^C - V_\pi^C(s_t) \right]^2, \tag{35}$$

where $G_t^R$ and $G_t^C$ are the discounted reward and cost in the buffer $D_{batch}$; $V_\pi^R(s_t)$ and $V_\pi^C(s_t)$ are the output of the reward and cost critic network. Due to the simple form of both reward and cost functions in CMDP model, value functions can be accurately approximated using critic networks with a simple structure.

*B. Implementation Details*

*1) Neural network structures of Actor and Critic:* The actor network $\pi_\theta(s)$, with parameters $\theta$, receives the state vector $s$ of environment and generates a series of probability distribution density function for actions. In this study, the actor network outputs the mean values of the action $\mu$, while standard deviation $\sigma$ is directly included in the DNN parameters $\theta$. The mean $\mu$ and standard deviation $\sigma$ are employed by actor to randomly sample actions from the Gaussian distributions $\mathcal{N}(\mu, \sigma)$. The critic networks $V_\varphi^R(s)$ and $V_\phi^C(s)$, with parameters $\varphi$ and $\phi$, receive the state vector $s$ and generate value function estimation. The overall structures of the proposed actor and critic networks are illustrated in Fig. 5 and Fig. 6.

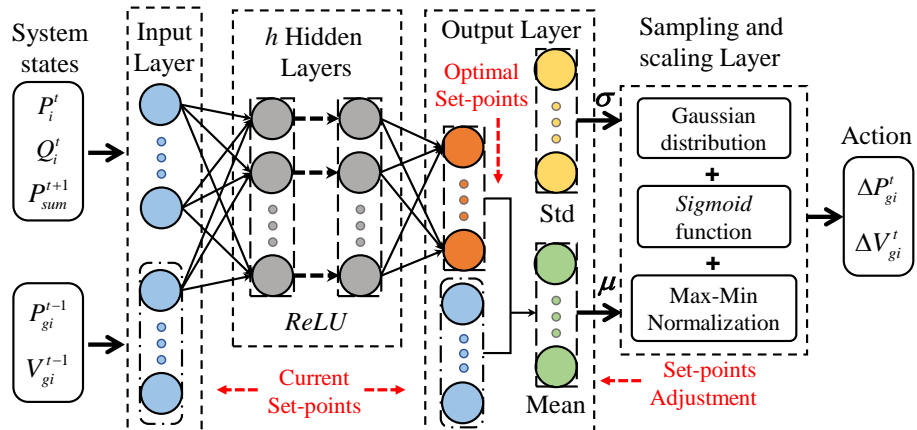

**Fig. 5.** *Structure of the proposed actor network.*

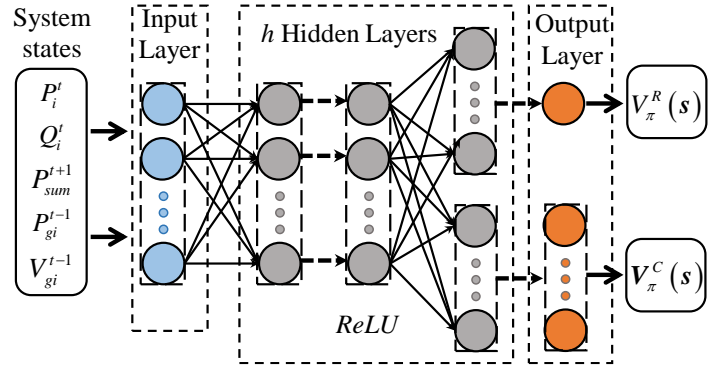

*Fig. 6. Structure of the proposed critic network.*

Before the state $s$ is fed into the DNN, the max-min normalization is utilized to normalize the system state into the interval $[0,1]$ to avoid excessive order-of-magnitude differences between features. The activation function of the output layer in the actor network is *sigmoid* function, which ensures the output value falls within the range of $[0,1]$ and its inverse normalization can definitely satisfy both active output and voltage constraints. Rectified Linear Unit (ReLU) is employed in the hidden layers to prevent "Vanishing Gradients".

The actor network decides the optimal set-points of generators. The output vector of DNN is defined as the mean values of the adjustments which are combined with the standard deviation to obtain a final random action policy. By sampling from the random policy, the action is determined after inverse max-min normalization and constraint checking.

$$\begin{cases} \Delta P_g^t = clip\left[\left(P_{g,\max} - P_{g,\min}\right) \cdot \mathcal{N}(\mu,\sigma), \Delta P_{g,\min}^t, \Delta P_{g,\max}^t\right] \\ \Delta V_g^t = clip\left[\left(V_{g,\max} - V_{g,\min}\right) \cdot \mathcal{N}(\mu,\sigma), \Delta V_{g,\min}^t, \Delta V_{g,\max}^t\right] \end{cases} \quad (36)$$

*2) Imitation Learning Implementation:* Random initialization for the actor network will result in large exploration space and long trial-and-error period, which affects the training efficiency negatively. Imitation learning (IL) can greatly shorten the stochastic exploration process of DRL agent by imitating the expert policies. Thus, IL methods are often used to pre-train actor networks before the DRL training.

As described in Section 3, the expert policy can be represented by an IPS. The actions of the expert can be calculated by running the IPS offline on training dataset. After obtaining the expert trajectories, interaction elements $(s_t \to a_t)$ can be reformulated as "feature" and "label". Behaviour cloning (BC) algorithms [37] can be employed to train the actor network to imitate the optimal generator set-points from the expert trajectories. The loss function is defined in (37).

$$L_\theta^{BC} = \frac{1}{N_{IL}} \sum_{(s_t,a_t)\in D_{IL}} \|a_t - \mu_\theta(s_t)\|_2^2, \quad (37)$$

where $D_{IL}$ is the training dataset of expert trajectories; $N_{IL}$ is the amount of training data in $D_{IL}$; $\mu_\theta(s_t)$ represents the mean action obtained by the actor network.

It is often insufficient to obtain the pre-trained policy for RT-OPF. Since most optimization problems have "binding" constraints, the optimal solution usually exists at the boundary of a high-dimensional feasible region. Due to the generalization error of DNN, the actions of IL agent may still violate the binding constraints and result in infeasible solutions. The significance of the following DRL training is exploring the neighbourhood of the pre-trained policy space and updating the DNN to obtain a secure and near-optimal policy. This process can be illustrated by Fig. 7 below.

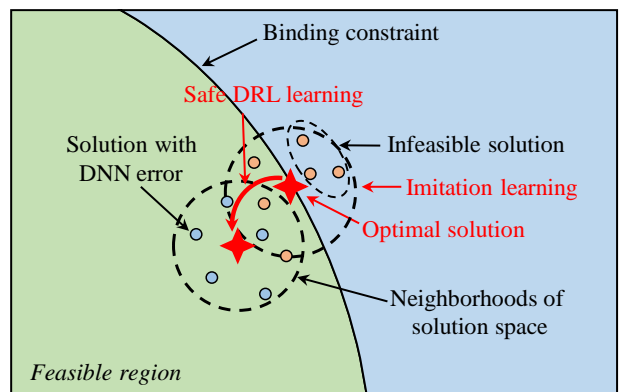

*Fig. 7. Cooperation relationship between IL and DRL.*

*3) Algorithm pseudocode:* The pseudocode of the PD-PPO algorithm for RT-OPF problem is summarized in Algorithm 1.

**Algorithm 1** The Pseudocode of the PD-PPO algorithm

| | |
|---|---|
| | Algorithm 1: PD-PPO for solving the RT-OPF problem |
| 1: | **Initialize:** the episode length $T$, the batch size $N_{batch}$, the training episode number $N_{episode}$, the training dataset $D_{train}$, parameters $\theta, \varphi, \phi$ of actor DNN and critic NN, discount factor $\gamma$ and $\lambda_{GAE}$, Lagrangian multipliers $\lambda_i$, clip factor $\varepsilon$, KL-divergence limit $KL_{tar}$, learning rate $\eta_\theta, \eta_\varphi, \eta_\phi, \eta_\lambda$, policy and critic updates number $N_\pi$ and $N_V$ |
| 2: | **For** each episode **do** |
| 3: |    Reset the environment state by sampling conditions from training dataset $D_{train}$. |
| 4: |    **For** each time step **do** |
| 5: |      Agent interacts with the environment and records interaction trajectory $\langle s_t, a_t, r_t, c_t, s_{t+1} \rangle$ as Fig. 3. |
| 6: |    **End For** |
| 7: |    **If** the replay buffer has been filled **do** |
| 8: |      Calculate $G_t^R$ and $G_t^C$ using (11) and (12). |
| 9: |      Calculate $A_{R,t}^{GAE}$, $A_{C,t}^{GAE}$, and $A_L^{\theta'}$ using (26)-(29) based on the critic network. |
| 10: |      **For** each gradient-updating step **do** |
| 11: |        Update actor network $\theta$ through optimizing (32). |
| 12: |        Update dual multipliers $\lambda$ using (33). |
| 13: |        **Break If** $D_{KL} > KL_{tar}$ |
| 14: |      **End For** |
| 15: |      **For** each gradient-updating step **do** |
| 16: |        Update critic networks $\varphi, \phi$ using (34) and (35). |
| 17: |      **End For** |
| 18: |      Empty replay buffer. |
| 19: |    **End If** |
| 20: | **End For** |
| 21: | **Return:** parameters of actor and critic networks $\theta, \varphi, \phi$. |

## 5. Case Studies

*A. Simulation settings*

The proposed PD-PPO algorithm is examined to solve the real-time AC OPF problem on the IEEE 9-bus, 30-bus and 118-bus transmission systems. Table I shows their partial characteristics, and the detailed parameters are obtained from the PYPOWER v5.1.16. In this experimental setup, Python v3.8.13 with Pytorch v1.12.1 is utilized as the programming environment, and the IPS included in PYPOWER is called to solve the RT-OPF problem as an expert policy. The key hyper-parameters of the proposed PD-PPO algorithm are presented in Table 2 below. All experiments are conducted on a desktop computer with an Intel i7-8700 CPU at 3.2 GHz and 16 GB RAM.

**Table 1** Test system characteristics

| Power system | $|\mathcal{N}|$ | $|\mathcal{L}|$ | $|\mathcal{G}|$ | $|\mathcal{S}|$ | $|\mathcal{A}|$ |
|---|---|---|---|---|---|
| 9-bus | 9 | 9 | 3 | 13 | 6 |
| 30-bus | 30 | 41 | 6 | 53 | 12 |
| 118-bus | 118 | 186 | 54 | 298 | 108 |

**Table 2** Hyper-parameters of the PD-PPO algorithm

| Parameter | Value | Parameter | Value |
|---|---|---|---|
| Optimizer | Adam | Discount factor | 0.95 |
| DNN types | Feedforward | Batch size | 32 |
| Activation function | ReLU | Replay buffer size | 400 |
| Actor learning rate | 5e-5 | $\lambda_{GAE}$ | 0.95 |
| Critic learning rate | 1e-4 | $KL_{tar}$ | 0.01 |
| $\lambda$ learning rate | 1e-3 | Clipping range $\varepsilon$ | 0.2 |

The net load curve is generated according to (21)-(24). In the following simulation experiments, the sampling range is set to 70% and 130% of the base values that can be obtained from the aforementioned PYPOWER for base demands at each bus. For the power factor in (23), the sampling range for power factor is $\left[0.9\beta_0, \min(1.1\beta_0, 1)\right]$, where $\beta_0$ is calculated by the base cases in PYPOWER. An example for curve generation process is shown in Fig. 8 for IEEE 30-bus system, where sub-figures display corresponding profiles of power demand, renewable output and net load at each bus. The historical data is from CAISO official website [38]. The dashed lines in Fig. 8 represent the historical data, while the solid lines illustrate the generated random demand data.

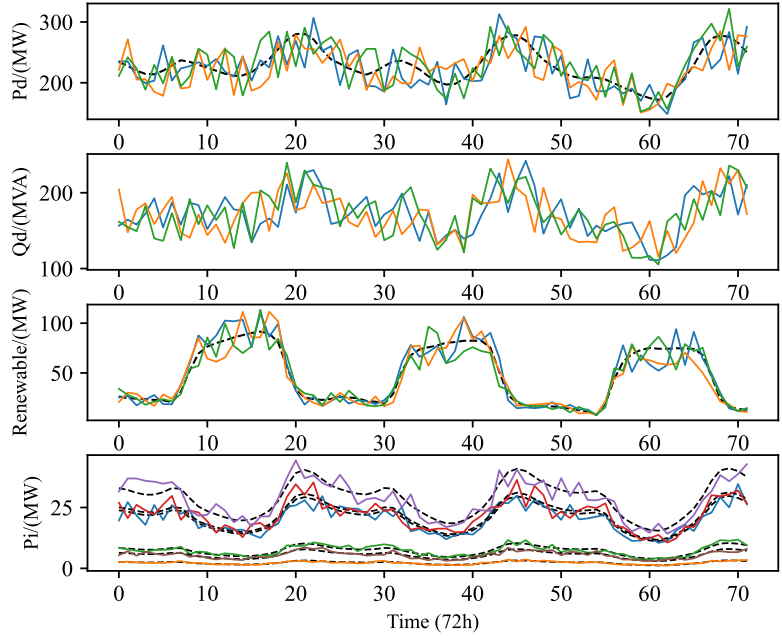

**Fig. 8.** *Randomness of the IEEE 30-bus test system.*

*B. Effectiveness of PD-PPO*

In this section, the performance of the proposed PD-PPO algorithm is compared with two other penalty-based PPO methods. Therefore, the methods for comparison include: i) Penalty-PPO introduces a penalty term into the reward function; ii) Cliff-PPO emphasizes the importance of the solution feasibility by constructing a "reward cliff", as illustrated in [27]. The reward function of Cliff-PPO method is as follows:

$$reward = \begin{cases} -\left(C_{pg} + C_{qg} + C_v + C_{flow}\right) & \text{Infeasible solution} \\ -kC_{generators} + b & \text{Feasible solution} \end{cases}, \quad (38)$$

where $C_{pg}, C_{qg}, C_v, C_{flow}$ are the violation of the security constraints; $C_{generators}$ is the operational cost of generators; $k$ and $b$ are both positive constants and $b$ is large enough to get a positive reward constantly. According to this reward function, the agent can receive a higher reward if and only if the solution is feasible. iii) the proposed PD-PPO method. The training performance for the three aforementioned methods is illustrated in Fig. 9 for the IEEE 30-bus and 118-bus systems.

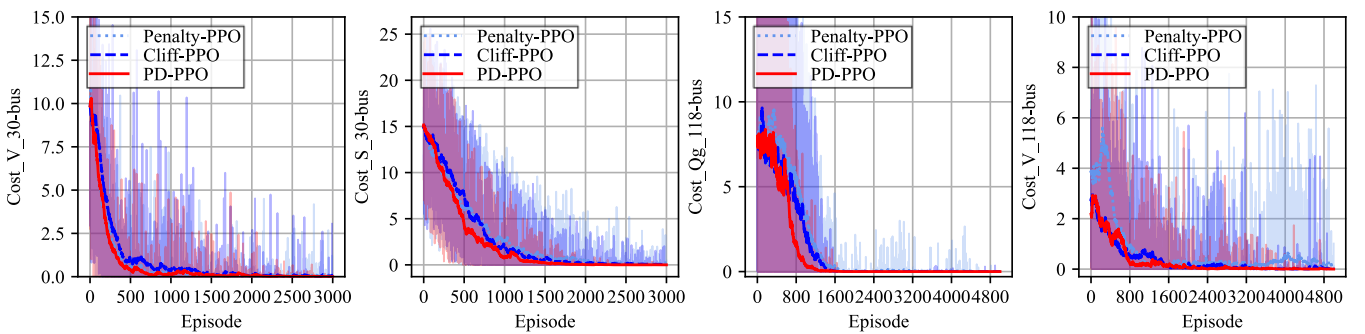

**Fig. 9.** *Training progress for IEEE 30-bus and IEEE 118-bus test system under three DRL methods.*

It is shown that the violation is larger during the early stages of the DRL training in Fig. 9. The reason is that some actions selected by IL agent will violate several security constraints. Owing to the efficiency of the PPO algorithm, three methods can effectively improve the security of the policy in both case studies. The Cliff-PPO offers faster learning process and lower cost than the Penalty-PPO due to the special design of the reward function. Because of the limitations of the MDP modelling, the policy security of the Penalty-based and Cliff-based PPO is difficult to guarantee. Finally, the proposed PD-PPO algorithm proves its ability for fast and stable learning, which has the fastest convergence rate and lowest cost at the end of the training process. The fundamental reason lies in the multi-dimensional evaluation perspective of the critic networks and the adaptive learning ability of the PD-PPO algorithm.

Fig. 10 illustrates the smoothed loss curve of the critic network for all three methods during the training process. The critic networks of the Penalty-PPO and Cliff-PPO exhibit slower decrease in loss, while the critic networks of the PD-PPO demonstrate rapid and stable convergence. The reward function with penalty terms will lead to increased complexity of the value function, which intensifies difficulty in the convergence of critic networks. Moreover, the critic estimation error usually leads to unstable actor network updates, further increasing the volatility of the policy performance.

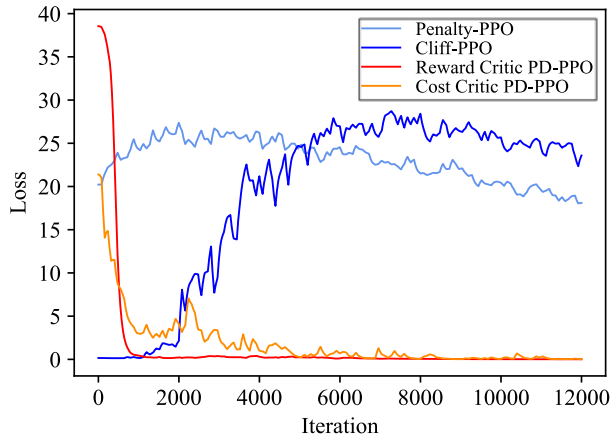

**Fig. 10.** *Critic network loss curve during the training process.*

The proposed PD-PPO algorithm employs two sets of critic networks to approximate the reward and cost separately, which reduces the complexity of the value function and improves critic estimation accuracy. Through updating the Lagrange multipliers, the PD-PPO agent could evaluate and balance the feasibility and optimality of current policy in a decoupled manner during the training process. For the above reasons, the proposed PD-PPO algorithm has more stable and efficient learning ability compared with the benchmark DRL algorithms.

*C. Operational constraint security*

The core objective of IL is to minimize the differences with actions from IPS rather than considering security constraints in the RT-OPF problem. The PD-PPO algorithm, as a safe DRL method, further updates the policy network pre-trained by IL to ensure the operational security.

The IEEE 30-bus test system is used as an example to facilitate the description of power flow, and its topology is depicted in Fig. 11. Fig. 12 compares the performance of the agent trained by IL and PD-PPO based on security constraint violations, while the states solutions obtained by IPS are also plotted for reference.

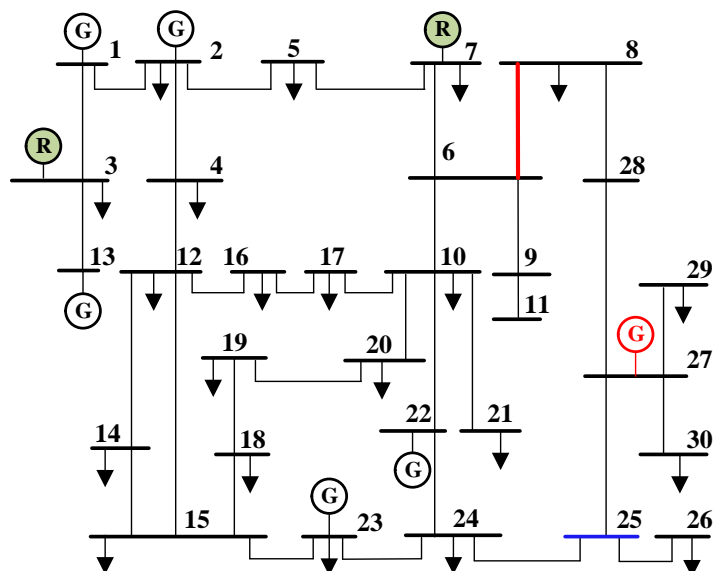

**Fig. 11.** *IEEE 30-bus test system topology.*

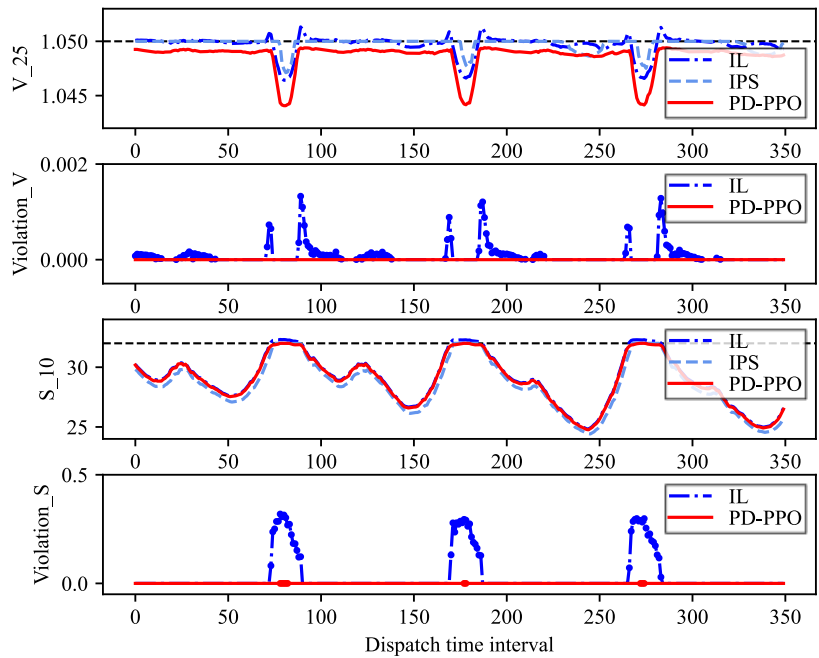

*Fig. 12. IL and PD-PPO performance for RT-OPF on the security constraints.*

From the solution obtained by IPS in Fig. 12, it can be seen that the voltage of bus 25 remains at 1.05 p.u. for the majority of time, while the transmission line connecting buses 6 and 8 operates at full capacity in high demand conditions. Therefore, these two sets of constraints are usually employed as "binding" constraints in the RT-OPF problem during the scheduling process. Fig. 12 shows that the PD-PPO agent could avoid violating the binding constraints with high security, while the non-negligible prediction errors of the IL agent led to violations of partial constraints. Due to the cooperative relation between the IL and safe DRL, the actor network initialized by IL is already very close to the optimal policy in the policy space, and the actions of PD-PPO agent and IL agent are relatively similar. Fig. 13 shows the IPS scheduling results and the PD-PPO agent decision trajectory, represented by dashed and solid lines, respectively.

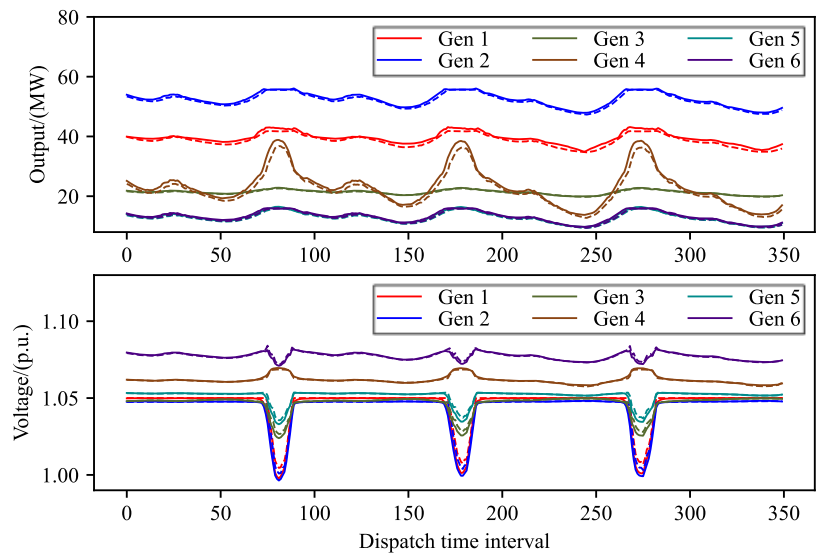

*Fig. 13. Comparisons of generator set-points for IPS and PD-PPO in IEEE 30-bus test system.*

For the IEEE 30-bus system, transmission line 6-8 is mainly used for transmitting power from the upper left area to the right area. Due to its limited capacity of only 32MVA, the transmission line often operates at full load during periods of high demand. Fig. 13 shows that the PD-PPO agent increases the output of the generator 4 to improve the active supply in right area, thus reducing the power flow across areas and ensuring the security of the transmission line constraint. Furthermore, in order to prevent overvoltage at bus 25, the $V_g$ obtained by the PD-PPO agent is slightly lower than the optimal solution from the IPS.

Two different contingency scenarios are configured to evaluate the performance of the proposed PD-PPO algorithm in response to contingencies. Fig. 14 shows the performance of the PD-PPO with expanded critic networks on contingency scenarios. N-1 contingency means outage of the transmission line which connects node 4-12, and N-2 contingency means outage of two transmission lines which connect nodes 4-12 and nodes 27-28. When the contingency occurs, the bus voltage variations of next day (24-48 steps) are shown in Fig. 14. After the contingency occurs, the voltage of a few buses will increase overall and violate the security constraints. Through the SDRL training for contingency scenarios, the PD-PPO agent can effectively avoid the overvoltage of the critical bus.

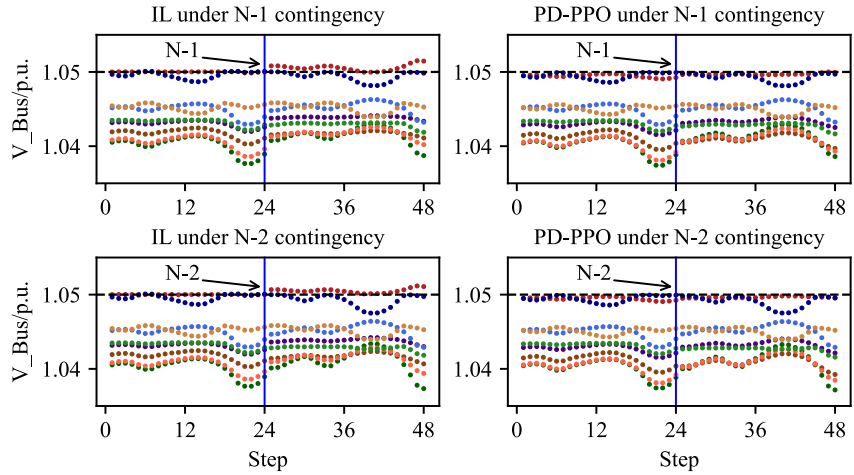

*Fig. 14. Nodal voltage variations of the power system after contingencies.*

Fig. 15 shows the cost function value of three DRL algorithms applied to the IEEE 30-bus and IEEE 118-bus system testing datasets. In this paper, we utilize solution success rate $Feas\%$ (40) and average cost $\bar{C}$ (41) to evaluate the performance of three DRL methods in terms of feasibility.

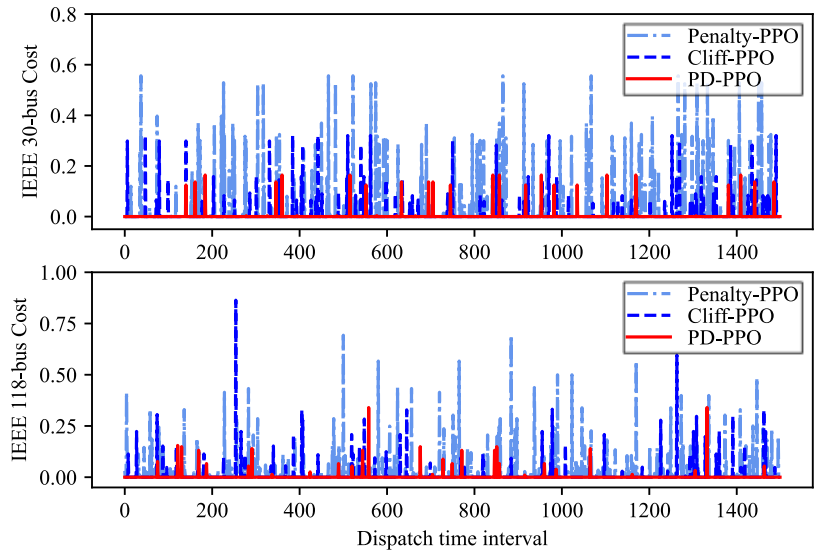

*Fig. 15. Comparisons of cost function on the testing dataset.*

$$feas_i = \begin{cases} 1 & \text{if } \left(\sum_{j=1}^{m}\left(C_j^{(i)} - d_j\right)\right) = 0 \\ 0 & \text{if } \left(\sum_{j=1}^{m}\left(C_j^{(i)} - d_j\right)\right) \neq 0 \end{cases}, i \in \{1,2,...,N_{Test}\}, \tag{39}$$

$$Feas\% = \frac{1}{N_{Test}} \sum_{i=1}^{N_{Test}} feas_i \times 100\%, \tag{40}$$

$$\bar{C} = \frac{1}{N_{Test}} \sum_{i=1}^{N_{Test}} \left[\sum_{j=1}^{m}\left(C_j^{(i)} - d_j\right)\right], \tag{41}$$

where $feas_i$ indicates whether the security constraints are all satisfied in the $i$th test scenario; $N_{Test}$ is the number of scenarios in the testing dataset. $Feas\%$ characterizes the probability of the successful solution. Then $\bar{C}$ denotes the average value of cost function obtained by different algorithms on the testing dataset.

**Table 3** Comparisons of the feasibility

| Power system | $Feas\%$ | | | $\bar{C}$ | | |
|---|---|---|---|---|---|---|
| | Penalty-PPO | Cliff-PPO | PD-PPO | Penalty-PPO | Cliff-PPO | PD-PPO |
| 9-bus | 97.10 | 100 | 100 | 0.0025 | 0 | 0 |
| 30-bus | 89.07 | 94.10 | 98.80 | 0.0243 | 0.0082 | 0.0023 |
| 118-bus | 83.33 | 95.80 | 97.93 | 0.0188 | 0.0068 | 0.0019 |

The results in Table 3 indicate that the PD-PPO method exhibits a higher solution success rate and lower constraint violations on the testing dataset compared to Penalty-PPO and Cliff-PPO. Penalty-PPO fails to ensure security of actions on testing dataset due to the limited impact of penalty terms for constraints. Cliff-PPO algorithm expects to satisfy the security constraints by constructing a reward "cliff", which performs better than the Penalty-PPO on feasibility. However, the effectiveness of the Cliff-PPO is limited by the low accuracy of critic network. The proposed PD-PPO algorithm has higher accuracy critic networks and hybrids the primal-dual method, which can fully consider the operational constraints and update to a more secure policy.

*D. Solution Optimality*

To verify the optimality of the proposed PD-PPO algorithm, the operational cost of the generators is compared with different OPF solution methods including IPS, IL, Penalty-PPO, and Cliff-PPO. We define $\kappa$ to measure the solution optimality relative to the objective function of the IPS.

$$\kappa = \frac{Cost_{agent} - Cost_{IPS}}{Cost_{IPS}} \times 100\% \tag{42}$$

Fig. 16 illustrates the optimality performance of the IPS, IL, and PD-PPO on test instances. The first subplot shows the operating cost curves of the three methods, which demonstrate a high degree of similarity. The other two subplots further calculate the $\kappa$ of IL and PD-PPO methods. Due to its ability to minimize the difference from the solver solution, the IL exhibits a very low $\kappa$, which is primarily concentrated below 0.01%. This result further indicates that the IL-initialized actor network can provide a high-quality initial policy for the DRL training. The $\kappa$ of the PD-PPO is slightly higher than IL, roughly concentrated at 0.5%. Based on the feasibility analysis, the PD-PPO is likely to converge towards a relatively conservative policy under the guidance of critic networks, which inevitably results in some loss of optimality. However, it is often adequate to find the near-optimal solution during the application, so the loss of optimality during training process is also acceptable.

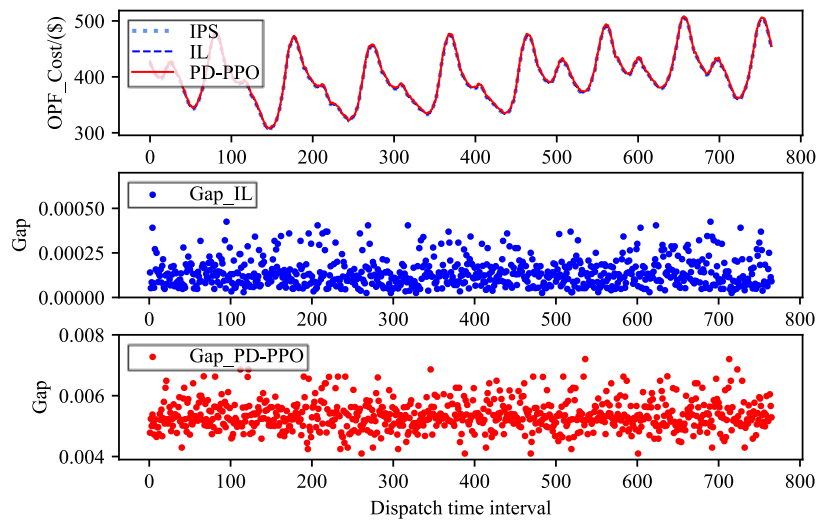

*Fig. 16. Comparisons of solution optimality on the testing dataset.*

Fig. 17 compares the overall performance of optimality for four AI methods on the testing dataset. Among the three types of PPO methods, Penalty-PPO has the lowest $\kappa$. The reason is that the penalty function serves as a "soft" constraint in the Penalty-PPO algorithm. As a result, the Penalty-PPO agent is more aggressive, resulting in lower loss of optimality but incomplete satisfaction of the security constraints. The Cliff-PPO agent has the largest $\kappa$. Actually, the Cliff-PPO is a special case of the Penalty-PPO, i.e., the Penalty-PPO algorithm with the infinite penalty coefficient. As a result, the policy of the Cliff-PPO may deviate significantly from the binding constraints and become more conservative, thereby affecting the near-optimality of the solution.

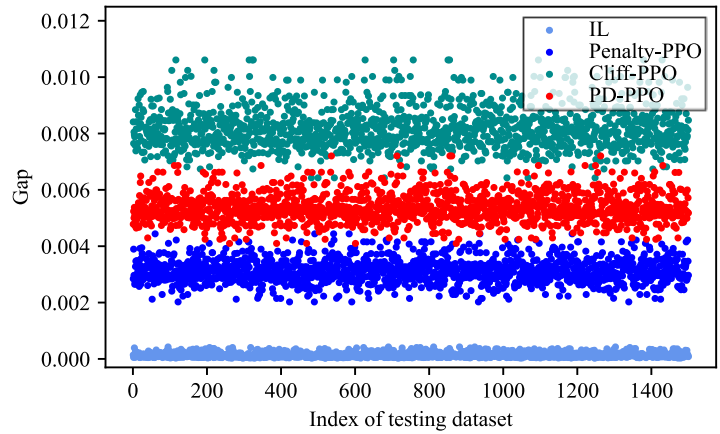

***Fig. 17.*** *The gap between the operating cost obtained by different methods and the solution of IPS.*

The maximum, minimum, and the average values of $\kappa$ for three types of PPO algorithms are compared in Table 4. It can be seen that the PD-PPO algorithm overcomes the disadvantages of these two types of algorithms and achieves a balanced effect. The PD-PPO algorithm alleviates the convergence difficulty of the critic network by mitigating reward sparsity through two sets of GAE critic system. Critic networks with high-precision estimation can improve the training efficiency while fully guaranteeing the security of converged policy. In addition, the primal-dual method enables adaptive coefficient adjustment during the training process, providing an alternative solution to the challenging issue of selecting appropriate penalty coefficients. The results show that the proposed method can guarantee the near-optimality of the policy as much as possible while satisfying the solution feasibility.

**Table 4** Comparisons of the optimality

| Power system | Penalty-PPO | | | Cliff-PPO | | | PD-PPO | | |
|---|---|---|---|---|---|---|---|---|---|
| | $\kappa_{max}$ (%) | $\kappa_{min}$ (%) | $\kappa_{aver}$ (%) | $\kappa_{max}$ (%) | $\kappa_{min}$ (%) | $\kappa_{aver}$ (%) | $\kappa_{max}$ (%) | $\kappa_{min}$ (%) | $\kappa_{aver}$ (%) |
| 9-bus | 0.008 | 0.003 | 0.007 | 0.067 | 0.014 | 0.039 | 0.052 | 0.009 | 0.035 |
| 30-bus | 0.443 | 0.202 | 0.311 | 1.061 | 0.643 | 0.818 | 0.721 | 0.409 | 0.537 |
| 118-bus | 1.202 | 0.044 | 0.403 | 1.312 | 0.063 | 0.707 | 1.571 | 0.058 | 0.674 |

*E. Solution acceleration*

For the PD-PPO, the calculation time includes two parts: agent decision-making and power flow calculation. The time of decision-making is actually the forward calculation time of the DNN which is related to the input dimensions. The PD-PPO alleviates the soaring computational pressure caused by increased dimension of the RT-OPF problems, and significantly reduces the computational complexity to achieve manageable decision-time. The power flow calculation is a nonlinear equation solution problem. Obviously, it takes much less time than the constrained optimization problem RT-OPF. Therefore, the proposed PD-PPO method could obtain the whole solution in the same time as solving nonlinear equations, thus achieving considerable acceleration of the computation. The acceleration effects on different scale test cases are summarized in Table 5.

**Table 5** Comparisons of the computation efficiency

| Power system | $T_{IPS}$ (s) | $T_{PF}$ (s) | $T_{Actor}$ (s) | *Speed up* |
|---|---|---|---|---|
| 9-bus | 0.39263 | 0.01192 | $2.843 \times 10^{-4}$ | ×27.37 |
| 30-bus | 0.46016 | 0.01354 | $2.996 \times 10^{-4}$ | ×33.24 |
| 118-bus | 0.98554 | 0.01820 | $3.053 \times 10^{-4}$ | ×53.26 |

**6. Conclusion**

This paper proposes a PD-PPO algorithm that combines the DRL algorithm with the primal-dual method to efficiently handle the security constraints in real-time AC OPF problems. By formulating the RT-OPF problem as a CMDP, this paper considers the security operational constraints independently. During the training process, two sets of critic and GAE estimation systems with similar structures are used to evaluate the reward and cost function. On this basis, a trade-off between optimality and feasibility is achieved by PD-PPO. Then, the agent is guided to regard feasibility as a prerequisite for optimality. The modeling and unique learning algorithms for security constraints make the learning process significantly more efficient. The case study on different

IEEE test systems demonstrates that the well-trained agent can effectively improve the computation speed while ensuring feasibility and optimality.

Future works include: i) considering the utilization of transfer learning or meta learning methods to improve the generalization of the proposed method for systems with varying topologies; ii) validating the effectiveness of the proposed approach by applying to more realistic systems with a wider range of uncertainties and more diverse control decisions; iii) exploring potential applications of safe DRL methods in other power system optimization problems.

**Acknowledgement**

This work is supported by the National Key R&D Program of China (2022YFB2403200).